# Variations in Orbital Elements of Planets

## S. I. Ipatov

*Keldysh Institute of Applied Mathematics, Russian Academy of Sciences, Miusskaya pl. 4, Moscow, 125047 Russia*
Received October 20, 1998

**Abstract**—Limits and characteristic periods of variations in orbital elements of planets were studied by numerical integration of equations of motion. Interrelations between the characteristic periods of variations in orbital elements of some planets were found.

## INTRODUCTION

Investigations of secular perturbations of orbital elements of planets were begun by Lagrange and continued by numerous researchers. Subbotin (1968, Chapter XVIII, section 7) reviewed these works. Applegate *et al.* (1986) and Nobili *et al.* (1989) studied the orbital evolution of five outer planets (from Jupiter to Pluto) over ± 100 and ±50 Myr, respectively. Applegate *et al.* (1986) presented the limits of variations of semimajor axes *a*, eccentricities *e*, and orbital inclinations *i* of the outer planets, as well as the plots of variations in orbital elements of Pluto. Nobili *et al.* (1989) studied the secular frequencies of the giant planets. Laskar (1988) integrated the secular equations for 8 planets on the interval of 30 Myr and presented the plots of secular variations in *e* and *i* for the terrestrial planets. A series of publications was devoted to the evolution of Pluto's orbit only (Milani *et al.*, 1989; Sussman and Wisdom, 1988; etc.). The libration period of the argument of Pluto's perihelion about 90° was found to be 3.78 Myr. The orbital inclination of this planet varies with periods of 34, 150, and 570 Myr in addition to the predominant variation with a period of 3.8 Myr. Laskar (1996) integrated averaged equations of motion on the interval (−10, 15 Gyr) and concluded that Mercury may collide with Venus within no more than 3.5 Gyr.

The problem of the orbital evolution of planets attracts our interest because we intend to develop approximate methods for the study of the orbital evolution of minor bodies under the influence of planets. For this purpose, the limits and the periods of characteristic variations of orbital elements should be known. The analytical estimates made previously are not sufficiently accurate, and the results of the numerical calculations reported in the literature give insufficient data for the development of these methods.

## RESULTS AND DISCUSSION

Ipatov and Hahn (1997a, b; 1999a, b) studied the orbital evolution of several minor bodies by numerical integration of the equations of motion for the system consisting of the Sun, 8 or 9 planets, and a body with zero mass. The orbital evolution on the time intervals of ± 0.1, ±1, and ±2 Myr was investigated with the use of the BULSTO integrator (Bulirsh and Stoer, 1966), and the orbital evolution on the intervals of ±20 and 200 Myr was studied with the use of the RMVS3 simplex integrator (Levison and Duncan, 1994). The RMVS3 integrator provides lower accuracy of the calculations, but the calculation time is one order of magnitude shorter. Using this integrator, we also obtained the time variations in the orbital elements shown in Fig. 1 mainly for the interval (±20, 0 Myr). On this time interval, the calculations were performed for 8 planets (except Mercury). Figure 1a shows the results of the calculations on the interval (−5, 0 Myr), and Fig. 1i illustrates the time dependences of *e* and *i* on the interval (0, 200 Myr). These dependences were constructed with the consideration of all 9 planets. With the exception of the plots for Pluto, the graphs on the interval of 200 Myr give no additional information as compared to the plots on the interval *T* = 20 Myr. The equations were integrated in the Cartesian system of coordinates. For *T* = 20 Myr, the calculations of the orbital elements were made with a step Δ*t* = 1000 yr. The values of *t* were determined relative to the initial (at *t* = 0) orbit of the Earth.

Table 1 gives the limits of the variations in semimajor axes, Δ*a* = *a*_max − *a*_min; minimum and maximum distances to the Sun (*R*_min, *R*_max); minimum and maximum values of eccentricity (*e*_min, *e*_max) and inclination (*i*_min, *i*_max), Δ*e* = *e*_max − *e*_min, and Δ*i* = *i*_max − *i*_min. With the exception of Mercury (*T* = 200 Myr), the data for all the planets are shown for the time interval (−20, 0 Myr).

Table 2 lists the average periods of variations in eccentricity (*T_e*), inclination (*T_i*), and the argument of the perihelion (*T_ω*) of the planet; in the difference in the longitude of ascending nodes of a planet and Jupiter, ΔΩ_J = Ω − Ω_J; and in the difference in the longitude of perihelions of a planet and Jupiter Δπ_J = π − π_J. These data, obtained by Ipatov (1997), are based on the analysis of the time dependences of the orbital elements. If the angles increase or decrease with time monotonically, then one period is the time it takes for the angle to change by 360°. Letters *I*, *D*, and *L* in the $k_\omega$, $k_{\Delta\Omega_J}$, and $k_{\Delta\pi_J}$ rows denote an increase, a decrease, and Il a libration of the angles, respectively. The notation signifies that the angle basically increases, but oscillates about zero for a certain amount of time; $t_I$ is the period of variations in $i$ exceeding $t_i$ (if such variations in $i$ are not well pronounced, the values of $t_I$ are not given in Table 2). With time, the periods can vary by several times. Table 2 gives the mean values of the periods, mainly for intervals of 1 and 20 Myr.

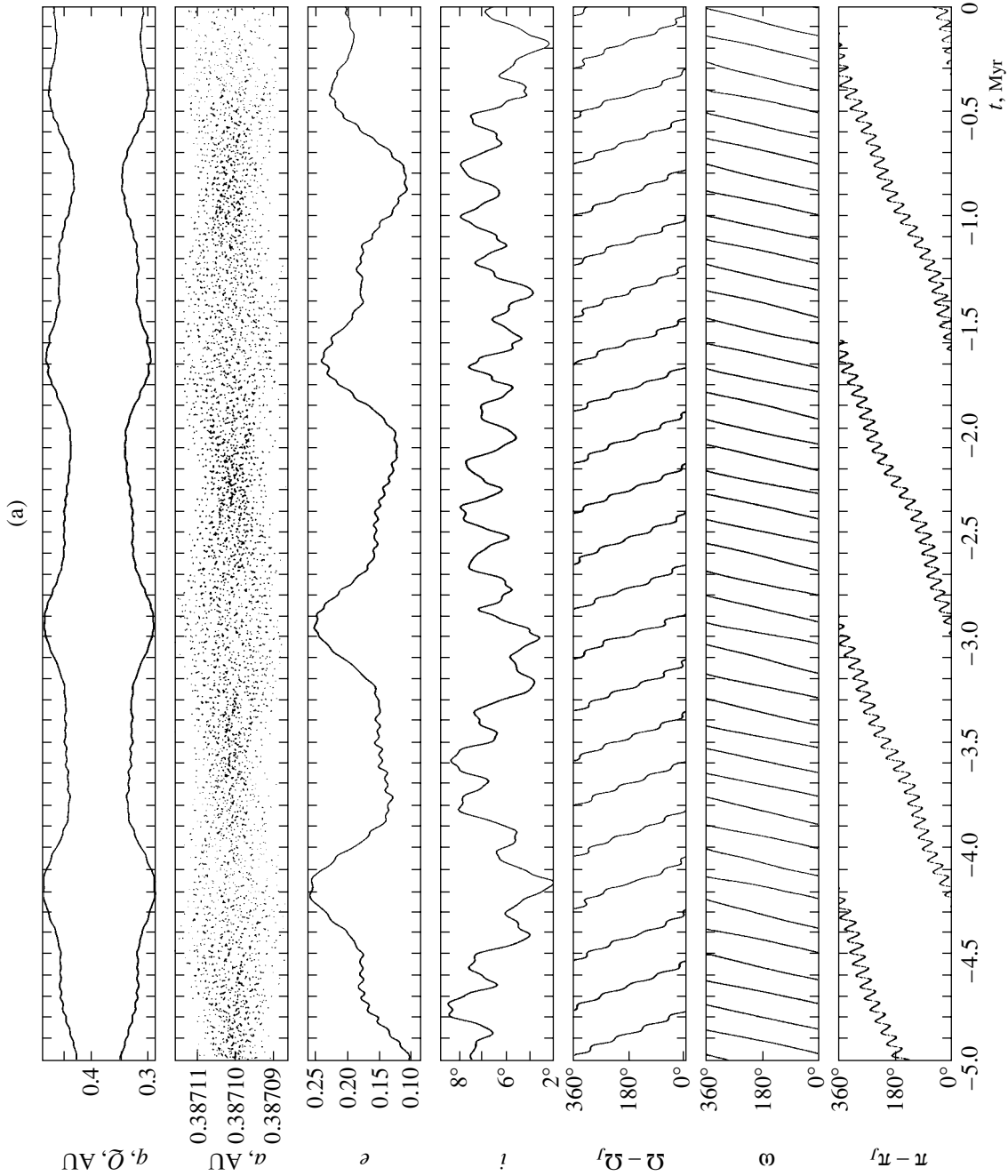

**Fig. 1.** Time variations (in Myr) in the semimajor axis $a$ (in AU); perihelion and aphelion distances, $q = a(1 - e)$ and $Q = a(1 + e)$ (in AU); eccentricity $e$; inclination $i$ to the initial orbital plane of the Earth (in deg); in the difference between longitudes of ascending nodes of a planet and Jupiter; $\Delta\Omega = \Omega - \Omega_J$; the argument of the perihelion $\omega$; in the difference between longitudes of the perihelions of a planet and Jupiter, $\Delta\pi = \pi - \pi_J$ (all angles are given in deg) for (a) Mercury, (b) Venus, (c) the Earth, (d) Mars, (e) Jupiter, (f) Saturn, (g) Uranus, (h) Neptune, and (i) Pluto. The results were obtained by numerical integration with the use of the RMVS3 integrator.

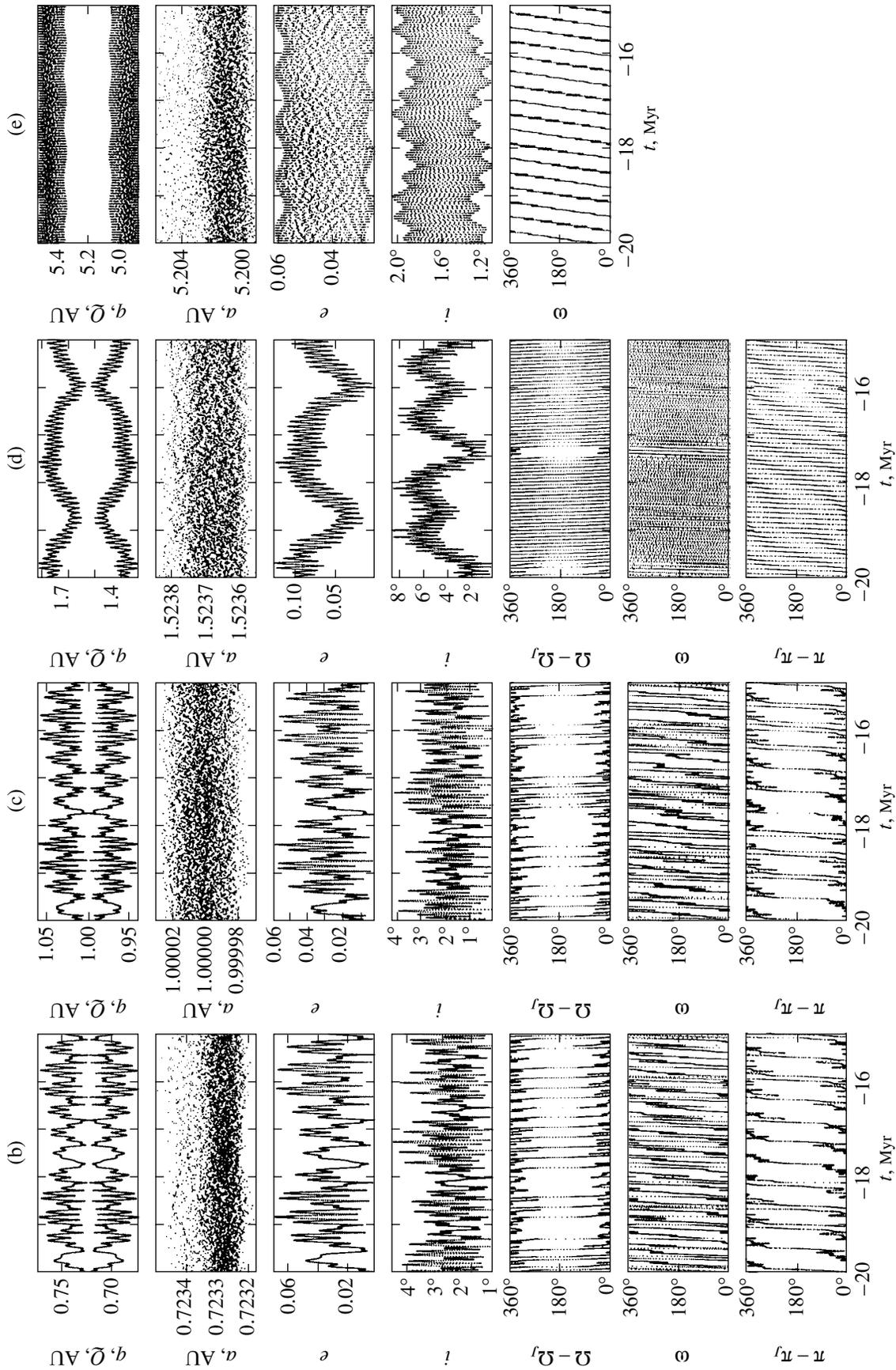

**Fig. 1.** (Contd.)

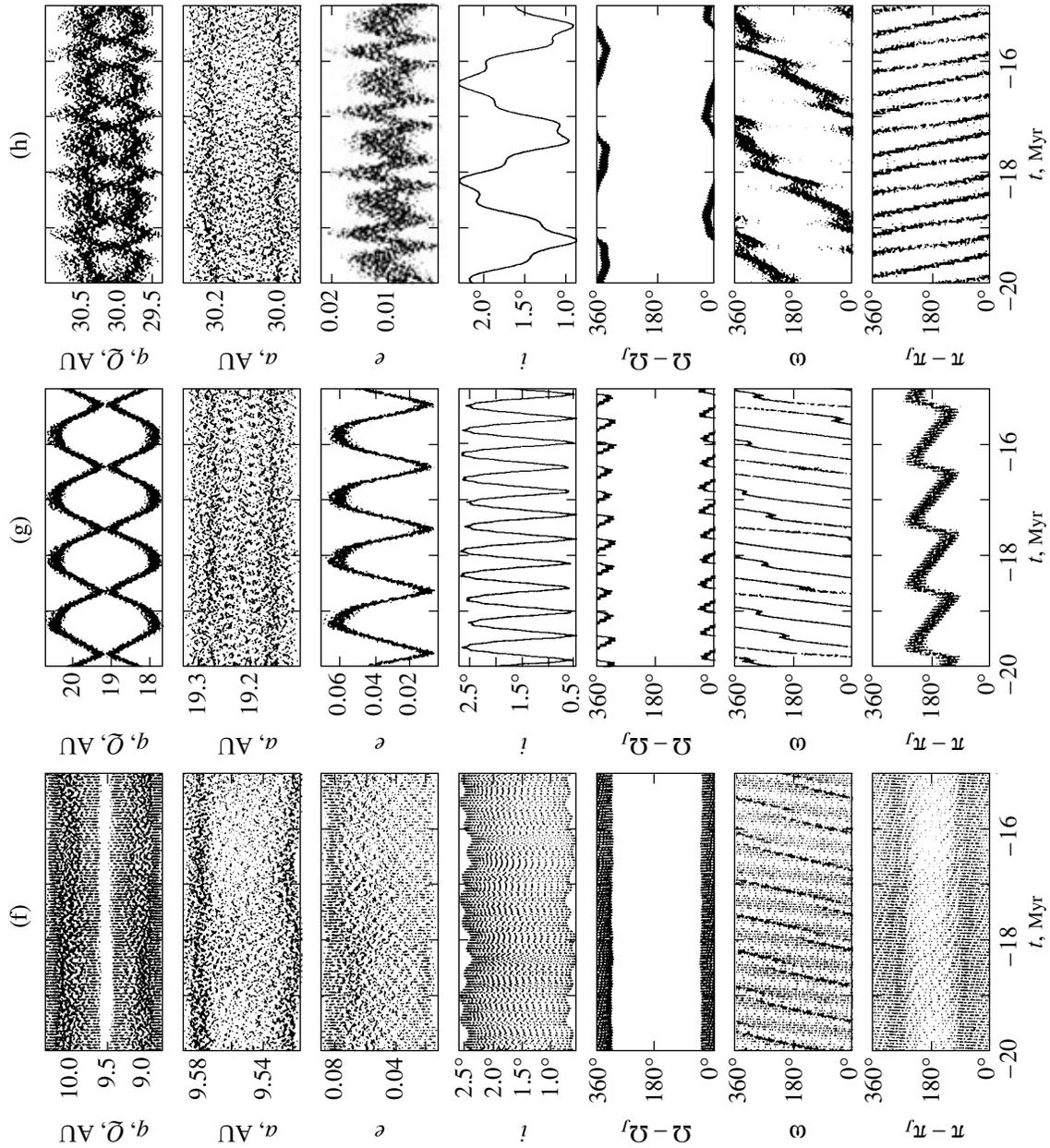

**Fig. 1.** (Contd.)

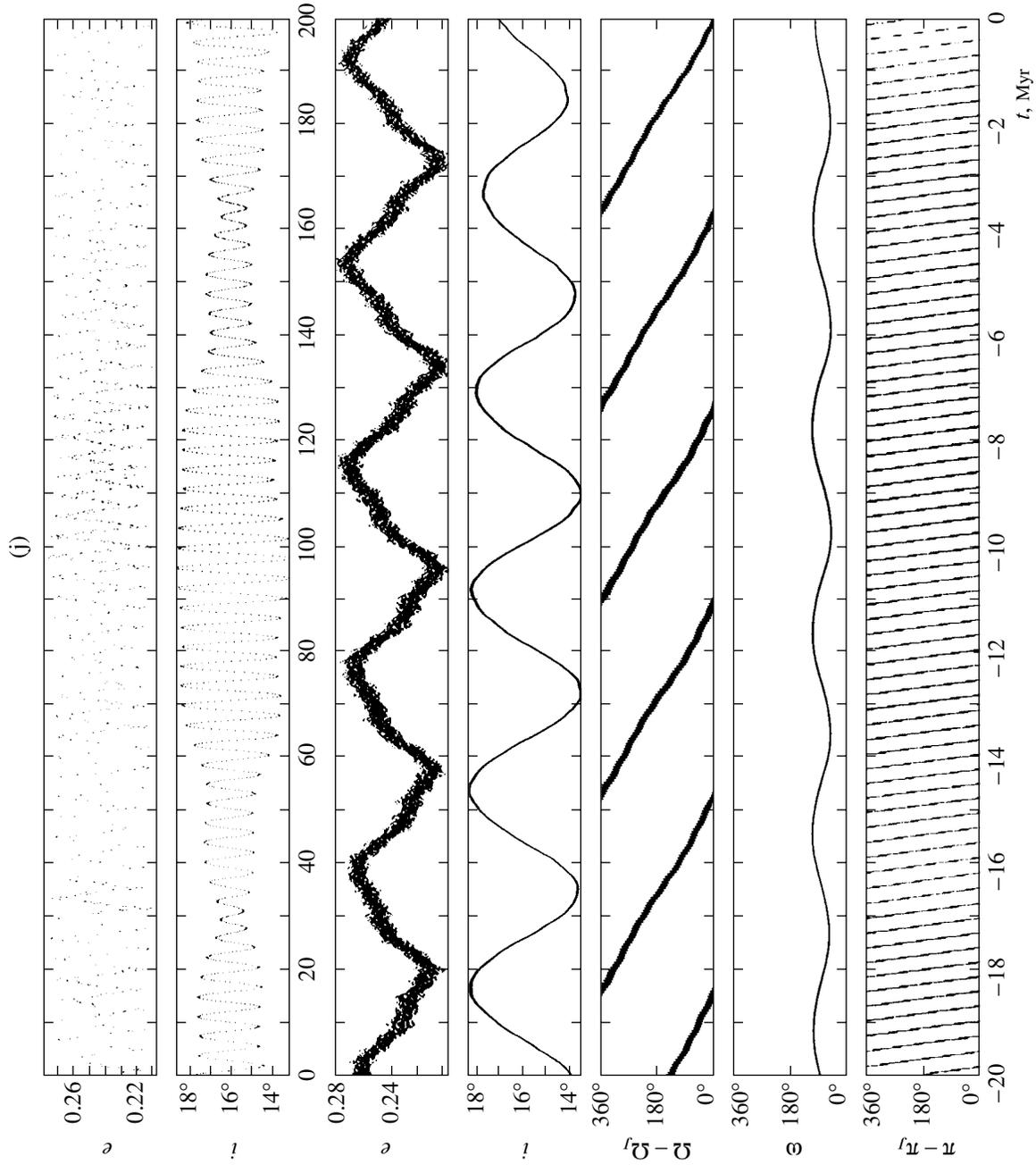

**Fig. 1.** (Contd.)

**Table 1.** Limits of variations of orbital elements

| Planet | Mercury | Venus | Earth | Mars | Jupiter | Saturn | Uranus | Neptune | Pluto |
|---|---|---|---|---|---|---|---|---|---|
| $\Delta a$, AU | 0.000030 | 0.000033 | 0.000062 | 0.00029 | 0.0037 | 0.080 | 0.231 | 0.407 | 1.004 |
| $R_{min}$, AU | 0.281 | 0.673 | 0.939 | 1.330 | 4.880 | 8.69 | 17.70 | 29.34 | 28.39 |
| $R_{max}$, AU | 0.493 | 0.774 | 1.061 | 1.717 | 5.524 | 10.45 | 20.78 | 31.00 | 51.15 |
| $e_{min}$ | 0.080 | 0.0002 | 0.0002 | 0.0018 | 0.0252 | 0.0074 | 0.0009 | 0.00005 | 0.2065 |
| $e_{max}$ | 0.273 | 0.0697 | 0.0608 | 0.1268 | 0.0618 | 0.0894 | 0.0757 | 0.0222 | 0.2806 |
| $\Delta e$ | 0.193 | 0.0695 | 0.0606 | 0.125 | 0.0366 | 0.082 | 0.0748 | 0.0222 | 0.0741 |
| $i_{min}$, deg | 1.80 | 0.013 | 0.000 | 0.061 | 1.096 | 0.567 | 0.438 | 0.783 | 13.414 |
| $i_{max}$, deg | 11.86 | 4.731 | 0.429 | 8.640 | 2.063 | 2.594 | 2.710 | 2.365 | 18.446 |
| $\Delta i$, deg | 10.06 | 4.718 | 0.429 | 8.579 | 0.967 | 2.027 | 2.272 | 1.582 | 5.032 |

**Table 2.** Periods of variations of orbital elements (in Myr)

| Planet | Mercury | Venus | Earth | Mars | Jupiter | Saturn | Uranus | Neptune | Pluto |
|---|---|---|---|---|---|---|---|---|---|
| $T_\omega$ | 0.12 | 0.1 | 0.1 | 0.036 | 0.30 | 0.046 | 0.30 | 1.91 | 3.8 |
| $k_\omega$ | *I* | *I* | *I* | *I* | *I* | *I* | *I* | *I* | *L* |
| $T_{\Delta\Omega_J}$ | 0.2 | 0.07 | 0.07 | 0.07 | – | 0.049 | 0.44 | 1.87 | 3.7 |
| $k_{\Delta\Omega_J}$ | *D* | *L, D* | *L, D* | *D* | – | *L* | *L* | *L* | *D* |
| $T_i$ | 0.2 | 0.07 | 0.07 | 0.07 | 0.049 | 0.049 | 0.43 | 1.87 | 3.7 |
| $T_e$ | 1 | 0.1 | 0.1 | 0.1 | 0.054 | 0.054 | 1.12 | 0.53 | 3.8 |
| $T_{\Delta\pi_J}$ | 1.4 | 0.3 | 0.2 | 0.1 | – | 0.054 | 0.054 | 0.36 | 0.28 |
| $k_{\Delta\pi_J}$ | *I* | *Il* | *Il* | *I* | – | *I* | *L* | *D* | *D* |
| $T_I$ | | 0.25 | 0.34 | 0.25 | 1.9 | 1.9 | | | 130 |

Although all the data shown in Fig. 1 and Tables 1 and 2 are obtained with the use of the simplex integrator, and the orbital elements are calculated with a rather large step, the values of $\Delta a = a_{max} - a_{min}$, $e_{min}$, and $e_{max}$ given in Table 1 either coincide with those reported by Applegate *et al.* (1986) for the outer planets or $T = 214$ Myr or differ from them in the last digit, i.e., the simplex integrator yields a rather high accuracy of the calculations.

The periods of the variations in the orbital elements are generally close to those reported by other authors for the giant planets. However, there is a distinction. In the work by Nobili *et al.* (1989), the secular frequency $g_7$ corresponds to the period $T_\omega \approx 0.42$ Myr. From Fig. 1g we see that the argument of Uranus' perihelion increases by 360° ~66 times over 20 Myr, i.e., $T_\omega \approx 0.3$ Myr.

Since the calculations were performed during my brief visit to the Berlin Institute for Planetary Research in 1996 with the aim to study the orbital evolution of Jupiter-approaching objects, the graphs were constructed only for $\Delta\Omega_J = \Omega - \Omega_J$ and $\Delta\pi_J = \pi - \pi_J$ (as far as we know, analogous plots were not considered by other authors), and the files with the results of the calculations were deleted immediately after the visit. Nevertheless, these graphs are also of interest, because they characterize the influence of Jupiter on the orbital evolution of other planets. For example, the values of $T_{\Delta\pi_J}$ for Saturn and Uranus are equal to the periods $T_e$ of variations in the eccentricity of orbits for Jupiter and Saturn (0.054 Myr). The $\pi_J$ value increases by 360° over 0.3 Myr. We found that $T_{\Delta\Omega_J} \approx T_i$ for all the planets (except Jupiter). The periods $T_{\Delta\Omega_J}$ for Venus, the Earth, and Mars are about the same (0.07 Myr). The values of $\Delta\Omega_J$ for Saturn, Uranus, and Neptune librate about zero, and the value of $\Delta\pi_J$ for Uranus librates about 180°.

## CONCLUSIONS

The limits and the characteristic periods of variations in the orbital elements of all the planets were obtained by numerical integration of the equations of motion on the time interval of up to 200 Myr. In particular, it was found that the values of $\Delta\pi_J$ for Saturn and Uranus are equal to $T_e$ for Jupiter and Saturn (54000 yr). For all the planets (except Jupiter), $T_{\Delta\Omega_J}$ is approximately equal to $T_i$.

## ACKNOWLEDGMENTS

This work was supported by the Russian Foundation for Basic Research, project no. 96-02-17892, and the "Astronomiya" Federal Science and Technology Program, Section 1.9.4.1.